\documentclass{article}

\PassOptionsToPackage{numbers, sort&compress}{natbib}
\usepackage[dblblindworkshop, final]{neurips_2026}
\workshoptitle{The 6th Workshop on Mathematical Reasoning and AI}

\usepackage[utf8]{inputenc}
\usepackage[T1]{fontenc}
\usepackage{hyperref}
\usepackage{url}
\usepackage{booktabs}
\usepackage{amsfonts}
\usepackage{amsmath}
\usepackage{nicefrac}
\usepackage{microtype}
\usepackage{xcolor}
\usepackage{graphicx}
\usepackage{listings}
\usepackage{enumitem}
\usepackage[most]{tcolorbox}
\usepackage{float}

\definecolor{codebg}{RGB}{248,248,248}
\definecolor{codeframe}{RGB}{220,220,220}
\definecolor{commentgreen}{RGB}{50,120,80}

\newcommand{\para}[1]{\noindent\textbf{#1}\hspace{0.5em}\ignorespaces}
\usepackage{tcolorbox}
\tcbuselibrary{listings}

\newcommand{\figwidth}{0.7}                 
\makeatletter
\renewcommand{\section}{\@startsection{section}{1}{\z@}{-1.0ex \@plus -0.3ex \@minus -0.2ex}{0.6ex \@plus 0.2ex}{\large\bf\raggedright}}
\makeatother
\AtBeginDocument{\setlength{\parskip}{4pt}          
  \setlength{\textfloatsep}{6pt}\setlength{\abovecaptionskip}{3pt}
  \setlength{\abovedisplayskip}{2pt}\setlength{\belowdisplayskip}{2pt}}
\newtcblisting{leaexample}{listing only,listing options={style=leatex},colback=codebg,colframe=codeframe,boxrule=0.5pt,arc=2pt,left=4pt,right=4pt,top=1pt,bottom=1pt,before skip=2pt,after skip=2pt}

\lstdefinestyle{leatex}{
    basicstyle=\ttfamily\footnotesize,
    backgroundcolor=\color{codebg},
    commentstyle=\color{commentgreen},
    breaklines=true,
    columns=fullflexible,
    keepspaces=true,
    showstringspaces=false,
    aboveskip=0pt,
    belowskip=0pt,
    escapeinside={<@}{@>},
}

\renewtcblisting{leaexample}{
    listing only,
    listing options={style=leatex},
    colback=codebg,
    colframe=codeframe,
    boxrule=0.5pt,
    arc=2pt,
    left=5pt,
    right=5pt,
    top=3pt,
    bottom=3pt,
}

\title{Lea: An Agent Backbone for Mathematician-Controlled Formalization in Lean}

\author{
  Shaswat Patel\textsuperscript{1,*}\thanks{Equal contribution.\\ Code available at:\url{https://github.com/VIDA-NYU/Lea}} \quad
  Daniel Aarao Reis Arturi\textsuperscript{1,*} \quad
  Kevin Gomez\textsuperscript{1,*} \\[0.5em]
  \textbf{Jaume de Dios Pont}\textsuperscript{1} \quad
  \textbf{Jorge Poco}\textsuperscript{2} \quad
  \textbf{Eduardo V. Teixeira}\textsuperscript{3} \\[0.5em]
  \textbf{Sam Westrick}\textsuperscript{1} \quad
  \textbf{Claudio Silva}\textsuperscript{1} \quad
  \textbf{Chinmay Hegde}\textsuperscript{1} \\[1em]
  \textsuperscript{1}New York University \quad
  \textsuperscript{2}Fundação Getulio Vargas \quad
  \textsuperscript{3}Oklahoma State University \\[0.5em]
  \texttt{\{spp9399,dar579\}@nyu.edu} \\[0.5em]
}

\begin{document}

\maketitle

\begin{abstract}
Agentic systems now formalize research-level mathematics in Lean, but they are built for autonomy: the mathematician states a target and inspects a result. The alternative, a general-purpose coding harness, is domain-neutral and must be assembled into a formalization workspace before any mathematics begins. Mathematicians, however, report wanting to keep high-level control over how an argument is decomposed and established. We present \textbf{Lea}, a Lean-specific agent backbone for mathematician-facing software rather than a single interface. Lea exposes runs, sessions, and typed events through one application-neutral API, and ships with Lean tooling (blueprints, LSP, Loogle, SafeVerify) already wired. We report two applications built on it: \textbf{LeaChat}, a web client for interactive proof authoring, and \textbf{LeaOverleaf}, which formalizes theorems annotated directly in an Overleaf document; a session begun in either continues in the other. In a case study, a mathematician outlined a proof of the Durfee square generating function in a few natural-language messages, and Lea produced a complete Lean development from Mathlib alone, following the intended decomposition closely and requiring substantive correction at a single point. Lea and both applications are released as open source so that further mathematician-facing tools can be built on the same backbone.
\end{abstract}
\section{Introduction}

Agentic systems have begun to formalize research-level mathematics in Lean~\citep{demoura2021lean4}. Two families dominate. \emph{Autonomous provers}~\citep{hubert2025,xin2024,ren2025,chen2025,achim2025,wang2025kimina,lin2025goedel} search proof space with domain-specific training and now reach olympiad level. \emph{Coding-agent harnesses} wrap a general-purpose agent such as Claude Code or Codex with Lean tools, skills, and MCP servers: Numina-Lean-Agent~\citep{liu2026} pairs Claude Code with an MCP server exposing the Lean LSP and Mathlib retrieval and solved all twelve Putnam~2025 problems; LeanMarathon~\citep{zhang2026leanmarathon} coordinates four Codex agents over an evolving proof DAG under CI checks and proved 258 lemmas across seven research-level targets. Both families show that long-horizon formalization is within reach, but in both the unit of interaction is an autonomous run. LeanMarathon confines human involvement to redesigning the harness between runs; Numina-Lean-Agent lets an expert revise a Markdown blueprint between invocations. The mathematician sits at the endpoints of the process.

That is not how mathematicians say they want to work. In a recent mixed-methods study of formalization workflows~\citep{collins2026}, 81.7\% of surveyed students and researchers preferred full or partial human control over formalization, and 66.7\% wished to retain high-level strategic control while delegating mechanical work; six of seven participants in the accompanying user study combined several AI tools rather than relying on one. Earlier work found that mathematical correctness and perceived helpfulness diverge when mathematicians interact with language models, motivating systems in which users can inspect and correct model behavior~\citep{collins2024}. The autonomy-first systems report the consequences themselves: Numina-Lean-Agent finds its own output overly result-oriented and in need of expert simplification~\citep{liu2026}, and LeanMarathon reports a run that, with the required Mathlib theory absent, produced an artifact passing every structural check while establishing nothing mathematically~\citep{zhang2026leanmarathon}.

\begin{figure}[t]
  \centering
  \includegraphics[width=\figwidth\textwidth]{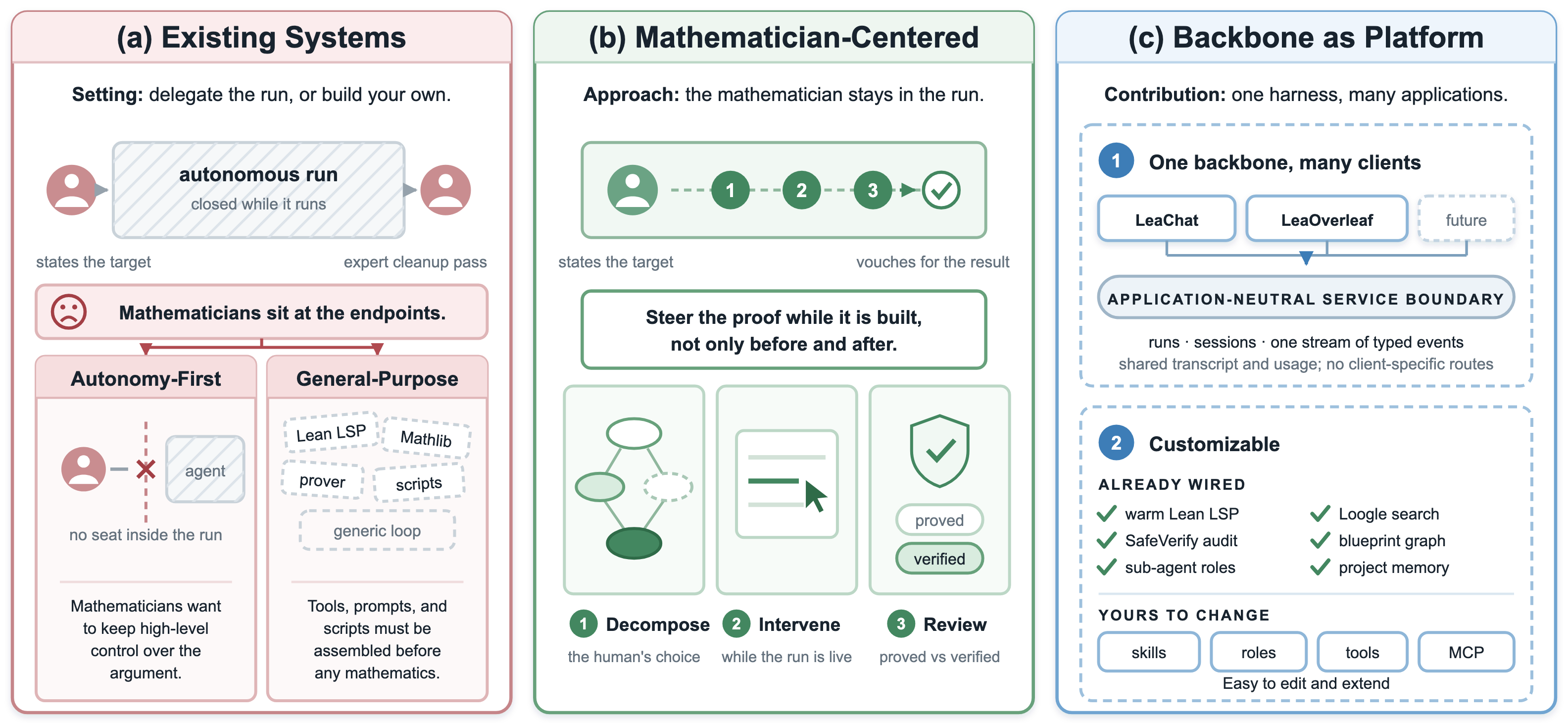}
  \caption{(a) Existing systems are \emph{autonomy-first} provers or \emph{general-purpose} harnesses. (b) Lea makes the run itself the surface the mathematician acts on. (c) One application-neutral service boundary carries every client; Lean tooling ships wired, while skills, sub-agent roles, tools, and MCP servers remain user-editable.}
  \label{fig:motivation}
\end{figure}

We present \textbf{Lea}, a Lean-specific agent backbone for mathematician-facing software (Figure~\ref{fig:motivation}). Relative to the systems above, Lea differs in four ways: the mathematician can intervene while a run is in progress rather than only between runs; the project blueprint persists across runs and is exposed as a dependency graph; Lean tooling (LSP, Loogle, blueprints, and kernel-replay verification with SafeVerify~\citep{safeverify}) ships wired rather than assembled per project; and the agent and project state are reachable through one application-neutral API, so several front ends can share a formalization. Our contributions are:
\begin{itemize}[leftmargin=*,itemsep=1pt,topsep=2pt]
  \item \textbf{An agent backbone} (\S\ref{sec:system}). One API exposes runs, sessions, and typed events, with no client-specific route. Multi-agent orchestration, persistent blueprints, and the Lean tools that recent harnesses integrated separately for benchmarking~\citep{liu2026,zhang2026leanmarathon,zhu2026leanarchitect,freer2025skills} are part of the system, and skills, sub-agent roles, tools, and MCP servers are user-editable without touching agent code.
  \item \textbf{Two applications on that backbone} (\S\ref{sec:applications}). LeaChat is a web client for interactive proof authoring. LeaOverleaf turns annotations in a \LaTeX{} document into blueprint nodes, writes each node's formalization status back into the document, and fingerprints every formalized statement so that edits to the paper or to the Lean import graph invalidate exactly the affected results. The two share sessions, transcripts, and usage accounting.
  \item \textbf{A case study} (\S\ref{sec:casestudy}). From a natural-language outline of the Durfee square generating function, Lea produced a complete Lean proof across three sessions and 38 sub-agent invocations with 28 human messages, one of which required substantive correction.
\end{itemize}

\section{The Lea Backbone}\label{sec:system}

Lea is a single backend for running and managing formalization agents, exposed through one HTTP API shared by all clients. A run involves three components: the \emph{formalization agent} decides what to do next, the \emph{Lean execution environment} carries that action out in the target project, and the \emph{persistent run history} records what happened and supplies it to later actions. The API exposes runs, sessions, and typed events rather than low-level tool calls, so an application can react to a proof file changing, a Lean check completing, or a sub-agent finishing without implementing agent logic of its own. Application state is derived from recorded Lean results rather than stored separately: whether a blueprint node is complete is read off the latest verification verdict for its proof.

\para{Formalization agent.}
The agent operates on a \emph{project}, a workspace that persists across runs and holds a Lean namespace, project instructions, accumulated memory, and a blueprint that decomposes the formalization into dependent lemmas~\citep{massot-blueprint}. Each blueprint node records its dependencies and the sessions that contributed to it; a node becomes \emph{ready} once its dependencies are complete. The agent can delegate bounded tasks to concurrent sub-agents, each with its own scratch directory, turn budget, and allowed tools. Roles are declared in YAML files that specify prompt, tools, budget, and optionally the model, so new roles are added without modifying the agent (Appendix~\ref{app:backbone}).

\para{Lean execution environment.}
The environment executes the agent's actions and checks them with the project's pinned toolchain and dependencies, including Mathlib, either locally or in a container. A persistent Lean language-server process reuses elaboration state across edits rather than starting Lean afresh for every check, and returns the effects of each action as typed events: file changes, diagnostics, search results, and verification verdicts. It establishes two machine-checked guarantees. A proof is \emph{formalized} when its file elaborates without \texttt{sorry}, and \emph{verified} when it additionally passes SafeVerify's kernel replay, per-declaration type matching, and axiom whitelist~\citep{safeverify}. The second matters for AI-generated proofs, which may compile while relying on a non-whitelisted axiom or proving a modified statement.

\para{Persistent run history.}
Every action and result is emitted as a typed event: messages, proof edits, tool results, Lean verdicts, sub-agent activity, usage, and run completion. This follows the action--observation model of OpenHands~\citep{wang2025openhands}, but Lea also makes the event stream the contract between backbone and applications, so clients receive an explicit ``proof changed'' or ``check completed'' rather than interpreting prover-specific tool output. The history is stored in SQLite with content-addressed proof snapshots, so a restored session shows exactly the proof state the live client saw. Each code step also records whether the mathematician or the agent made the change, so their contributions can be measured directly from the run history.
\section{Applications Built on Lea}\label{sec:applications}

The backbone separates the formalization machinery from the way mathematicians interact with it: Lea owns the agent loop, Lean execution, verification, and run history, while each application decides how users start, guide, and review a formalization. LeaChat and LeaOverleaf share projects, sessions, proof history, and verification results through the same API, and a future application supports a new workflow by sending commands to that API and presenting its typed events differently. Interface illustrations of both are provided in Appendix~\ref{app:screenshots}.

\para{LeaChat.}
LeaChat is a web client for sustained, interactive proof development. It shows the run transcript beside an editable Lean canvas, so the mathematician follows proof changes and verification results as they occur. While the agent is running, the mathematician can edit the current proof in the canvas, approve or deny a tool call, rewrite a blueprint node, or redirect the strategy through chat; each intervention is recorded in the run history and supplied as context to subsequent actions, so it shapes the proof as it develops rather than serving as retrospective feedback. LeaChat also manages skills, tools, MCP servers, and sub-agent roles without touching the harness, preserves each sub-agent's trajectory for inspection, and can reopen a session created in LeaOverleaf and continue it.

\para{LeaOverleaf.}
A paper under development already fixes the statements, proposes an order of dependence, and surrounds each claim with prose recording its intent; extracting statements into a separate harness discards this and leaves the author maintaining two developments. LeaOverleaf therefore treats the Overleaf project as the surface from which formalization is directed, dividing labour as the survey evidence recommends~\citep{collins2026}: the mathematician chooses the objects and decomposition, the agent produces and checks Lean. Figure \ref{fig:lea_overleaf} illustrates the LeaOverleaf interface.

The author marks a definition or theorem with a non-rendering comment naming the target, its dependencies, and optional guidance. From these annotations LeaOverleaf assembles a blueprint in the sense of \S\ref{sec:system}, with ready, blocked, and complete targets, and supports batch formalization of an eligible subgraph in topological order. Each formalization is a bounded run through the shared backbone: \LaTeX{} sources are mirrored into the project as readable source material, declared dependencies resolve to their recorded Lean artifacts, and an unresolved dependency blocks the run rather than being dropped. When a source proof is supplied, the agent is instructed to preserve the statement’s meaning and the argument’s essential mathematical approach. Without a supplied proof, it may attempt a meaning-preserving proof of a precise statement while disclosing that no source method was available. Runs proceed without per-tool approval, while authors can pause execution and continue through each item’s persisted chat. Individual runs can additionally pause when source issues require author input, if this option is enabled.

Results return as statuses. \emph{Formalized} means the Lean elaborates without \texttt{sorry}, \emph{verified} means SafeVerify also passes, and \emph{disproved} is a successful terminal outcome in which counterexample search machine-checked a refutation, for an author revising a draft among the most informative results a run can produce. Even verified asserts only that the proof establishes the Lean statement, not that the statement expresses what the author intended, the \emph{faithfulness gap}~\citep{faithfulnessgap2026}. LeaOverleaf keeps this gap visible with a final status, \emph{audited}, that no machine process can set: a mark recorded by the mathematician after review, bound to a hash of the annotation, the Lean artifact, and every project-local file it transitively imports, and removed if any input changes. 
Live Lea status reports expose the formalizing agent’s assessment of correspondence with the source, the scope of that assessment, and findings such as additional assumptions or departures from the supplied argument. They distinguish Lean-side repairs from unresolved source concerns. These reports are self-assessments, not independent or calibrated faithfulness checks. Review of the resulting statement and dependencies remains necessary. Appendix Figure~\ref{fig:lea_overleaf_faithfulness} shows an example report alongside the Lean artifact it describes.

Because a paper changes continually, each artifact is bound to a fingerprint of its statement text, kind, dependencies, and guidance, and changing any of these marks the result stale. Lean-side changes propagate through the import graph: a revised definition rebuilds transitive dependents and reports which results survive (a successful edit confined to a theorem’s proof body, with its declaration unchanged, need not invalidate dependents, although review and verification evidence remain bound to artifact revisions), and broken dependents can be repaired by hand or delegated back to Lea as a topologically ordered batch. Appendix~\ref{app:toy} walks through a small example in which a run ends in \emph{disproved} and exposes a missing hypothesis in the paper rather than an error in the Lean.

LeaOverleaf is a Chrome extension plus a local companion process. The companion sends formalization and continuation requests to the same adapter used by LeaChat, which invokes the shared prover in-process. Source-bound prompting and live status reporting are supported by the shared backend, while document integration remains in the extension and companion.

\begin{figure}
  \centering
  \includegraphics[width=0.8\textwidth]{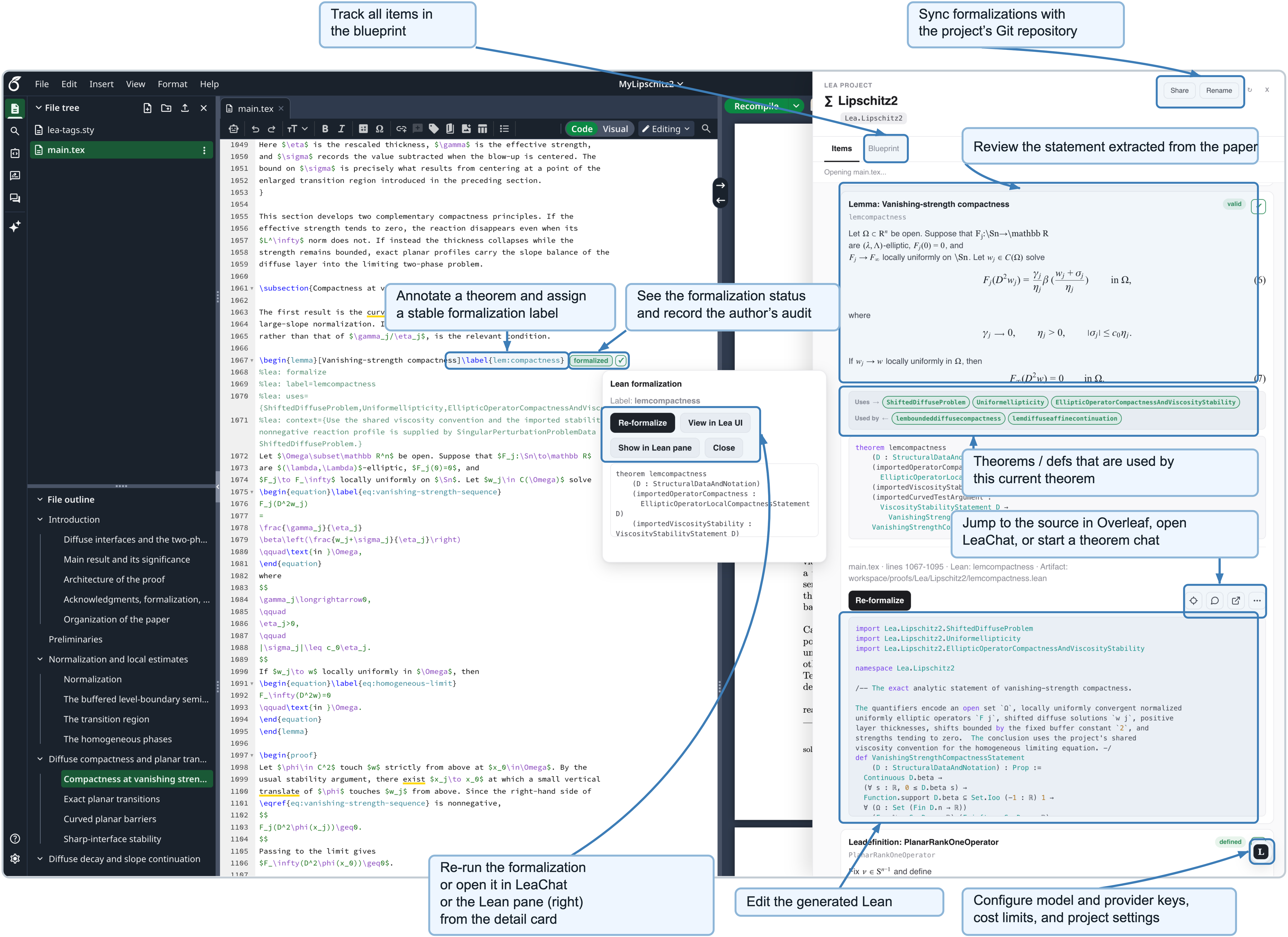}
  \caption{\textbf{The LeaOverleaf interface.} Left: an Overleaf document whose \LaTeX{} source contains Lea annotations that identify a theorem, dependencies, and audit status. Center: the item detail card, from which the author can rerun the formalization, continue it in LeaChat, or open it in the Lean pane. Right: the Lean pane, which organizes annotated items as a blueprint, displays the statement extracted from the paper and its dependencies, and presents the generated Lean for review and editing. Lean pane also provides additional controls such as navigate to the source in Overleaf,  synchronize with the Git repository, and configure the agentic harness.}
  \label{fig:lea_overleaf}
\end{figure}

\section{Case Study: The Durfee Square Generating Function}\label{sec:casestudy}

A \emph{partition} of $n$ is a non-increasing sequence of positive integers summing to $n$; $p(n)$ counts them. Euler's generating function $\sum_{n \ge 0} p(n)X^n = \prod_{k\ge1}(1-X^k)^{-1}$ is in Mathlib (since Lean 4.26) in a generalized form with a character function restricting part multiplicities, but many partition-theoretic results rest on combinatorial arguments on Ferrers diagrams that Mathlib does not model and that require elaborate Lean proofs~\citep{partitions}. The \emph{Durfee square} of $\lambda$ has side $s$, the largest $s$ with $\lambda_s \ge s$; removing it leaves a right residue with at most $s$ parts and a bottom residue with parts at most $s$, of combined size $n - s^2$, and conjugation shows the two families equinumerous. Counting over $s$ gives
\begin{equation}\label{eq:pn-durfee}
  \sum_{n=0}^{\infty} p(n)X^n = \sum_{s=0}^{\infty} \frac{X^{s^2}}{\prod_{j=1}^{s}(1-X^j)^2}.
\end{equation}
This outline suffices for a human reader; Lean requires the many discrete facts such a diagrammatic argument elides.

The mathematician gave this outline to a LeaChat project over several colloquial chat messages, with no PDF or \LaTeX{} source. Using GPT-5.5, Lea produced a Lean proof of (\ref{eq:pn-durfee}) from Mathlib alone, which supplies \texttt{Nat.Partition} and the generalized Euler theorem but no notion of Ferrers diagrams or Durfee squares. The proof spans three sessions and 38 sub-agent invocations, is organized by a blueprint worked out with the agent, and states the final theorem as
\begin{leaexample}
theorem genFun_eq_tsum_qPochhammerInvNat_sq :
    Nat.Partition.genFun = Σ' k : ℕ, PowerSeries.X ^ (k * k) *
        qPochhammerInvNat k * qPochhammerInvNat k := by
  ...
\end{leaexample}
The mathematician sent 28 messages in total, most of them confirmations of a definition or proof approach the agent proposed; every step introducing a new core definition received such confirmation. The mathematician deliberately stated little mathematical content, so that Lea's ability to make decisions and draw on Mathlib could be assessed. One sub-agent near the end required substantive correction, having proved a trivial restatement of an earlier lemma rather than the target. The resulting development follows the intended outline closely, an advantage over one-shot autonomous generation, though it still needs stylistic clean-up before it would meet Mathlib's standards (Appendix~\ref{app:casestudy}).
\section{Limitations and Conclusion}\label{sec:conclusion}

Our evidence is the architecture and one case study. Lea is also in use in ongoing research-level formalization projects, which we do not report here because their subject matter would identify the authors; the camera-ready version will describe them. We do not yet offer a controlled comparison with other harnesses such as Numina-Lean-Agent~\citep{liu2026} or an evaluation across mathematical domains, and Lea's output needs refactoring and expert review before it meets Mathlib's quality and disclosure requirements; nearer-term uses are Lean developments accompanying new papers, machine-checked proofs or refutations of LLM-generated conjectures, and formal components of AI-assisted projects such as TauCeti~\citep{tauceti}. Lea does not yet support several mathematicians working concurrently on one project, a natural next step for LeaOverleaf users who already share a document (Appendix~\ref{app:future}). What Lea does provide is formalization-specific infrastructure, with the mathematician in control during the run, that several applications can share without embedding a prover or reimplementing an agent loop, and on which further tools for mathematicians can be built.

\begin{ack}
This work was funded by the DARPA expMath program (Agreement No. HR0011262E029). Silva was also partially supported by NSF grants OAC-2411221 and DRL-2604563. Jorge was also partially supported by the Brazilian National Council for Scientific and Technological Development (CNPq), grant \# 313454/2026-4, and the Getulio Vargas Foundation (FGV). Any opinions, findings, conclusions, or recommendations expressed in this material are those of the authors and do not necessarily reflect the views of DARPA, NSF, or the authors' affiliated institutions.
\end{ack}

\bibliographystyle{plainnat}
\bibliography{refs_updated}

@inproceedings{demoura2021lean4,
  title     = {The {Lean} 4 Theorem Prover and Programming Language},
  author    = {de Moura, Leonardo and Ullrich, Sebastian},
  booktitle = {Automated Deduction -- {CADE} 28},
  series    = {Lecture Notes in Computer Science},
  volume    = {12699},
  pages     = {625--635},
  year      = {2021},
  publisher = {Springer}
}

@misc{massot-blueprint,
  title        = {leanblueprint: A Tool for Writing Blueprints of {Lean} 4 Projects},
  author       = {Massot, Patrick},
  howpublished = {\url{https://github.com/PatrickMassot/leanblueprint}},
  year         = {2024}
}

@misc{safeverify,
  title        = {{SafeVerify}: Verifying {Lean} 4 Proofs by Kernel Replay,
                  Per-Declaration Type Matching, and an Axiom Whitelist},
  author       = {{GasStationManager}},
  howpublished = {\url{https://github.com/GasStationManager/SafeVerify}},
  year         = {2025}
}

@article{axle,
  title   = {{AXLE}: A Cloud Infrastructure for {Lean} 4 Theorem Proving Utilities},
  author  = {Xin, Jimmy and Schneidman, Alex and Cummins, Chris and Ram, Karun and
             Ganesh, Srihari and Limperg, Jannis},
  journal = {arXiv preprint arXiv:2606.26442},
  year    = {2026}
}

@misc{freer2025skills,
  title        = {{Lean} 4 Skills: Theorem Proving Skill and Workflow Pack for
                  {AI} Coding Agents},
  author       = {Freer, Cameron},
  howpublished = {\url{https://github.com/cameronfreer/lean4-skills}},
  year         = {2025}
}

@article{zhu2026leanarchitect,
  title   = {{LeanArchitect}: Automating Blueprint Generation for Humans and {AI}},
  author  = {Zhu, Thomas and Monticone, Pietro and Avigad, Jeremy and Welleck, Sean},
  journal = {arXiv preprint arXiv:2601.22554},
  year    = {2026}
}

@article{hubert2025,
  title   = {Olympiad-Level Formal Mathematical Reasoning with Reinforcement Learning},
  author  = {Hubert, Thomas and Mehta, Rishi and Sartran, Laurent and others},
  journal = {Nature},
  year    = {2025},
  doi     = {10.1038/s41586-025-09833-y}
}

@article{xin2024,
  title   = {{DeepSeek-Prover-V1.5}: Harnessing Proof Assistant Feedback for
             Reinforcement Learning and {Monte-Carlo} Tree Search},
  author  = {Xin, Huajian and Ren, Z. Z. and Song, Junxiao and others},
  journal = {arXiv preprint arXiv:2408.08152},
  year    = {2024}
}

@article{ren2025,
  title   = {{DeepSeek-Prover-V2}: Advancing Formal Mathematical Reasoning via
             Reinforcement Learning for Subgoal Decomposition},
  author  = {Ren, Z. Z. and Shao, Zhihong and Song, Junxiao and others},
  journal = {arXiv preprint arXiv:2504.21801},
  year    = {2025}
}

@article{wang2025kimina,
  title   = {{Kimina-Prover} Preview: Towards Large Formal Reasoning Models with
             Reinforcement Learning},
  author  = {Wang, Haiming and Unsal, Mert and Lin, Xiaoran and Baksys, Mantas and
             Liu, Junqi and Dos Santos, Marco and others},
  journal = {arXiv preprint arXiv:2504.11354},
  year    = {2025}
}

@article{lin2025goedel,
  title   = {{Goedel-Prover-V2}: Scaling Formal Theorem Proving with Scaffolded
             Data Synthesis and Self-Correction},
  author  = {Lin, Yong and Tang, Shange and Lyu, Bohan and Yang, Ziran and
             Chung, Jui-Hui and Zhao, Haoyu and others},
  journal = {arXiv preprint arXiv:2508.03613},
  year    = {2025}
}

@article{achim2025,
  title   = {Aristotle: {IMO}-Level Automated Theorem Proving},
  author  = {Achim, Tudor and Best, Alex and Bietti, Alberto and others},
  journal = {arXiv preprint arXiv:2510.01346},
  year    = {2025}
}

@article{chen2025,
  title   = {{Seed-Prover} 1.5: Mastering Undergraduate-Level Theorem Proving via
             Learning from Experience},
  author  = {Chen, Jia and Chen, Wei and Du, Jing and others},
  journal = {arXiv preprint arXiv:2512.17260},
  year    = {2025}
}

@article{liu2026,
  title   = {{Numina-Lean-Agent}: An Open and General Agentic Reasoning System for
             Formal Mathematics},
  author  = {Liu, Junqi and Zhou, Zihao and Zhu, Zekai and others},
  journal = {arXiv preprint arXiv:2601.14027},
  year    = {2026}
}

@article{zhang2026leanmarathon,
  title   = {{LeanMarathon}: Toward Reliable {AI} Co-Mathematicians through
             Long-Horizon {Lean} Autoformalization},
  author  = {Zhang, Yuanhe and Sun, Yuekai and Suzuki, Taiji and Lee, Jason D. and
             Liu, Fanghui},
  journal = {arXiv preprint arXiv:2606.05400},
  year    = {2026}
}

@inproceedings{wang2025openhands,
  title     = {{OpenHands}: An Open Platform for {AI} Software Developers as
               Generalist Agents},
  author    = {Wang, Xingyao and Li, Boxuan and Song, Yufan and others},
  booktitle = {International Conference on Learning Representations ({ICLR})},
  year      = {2025}
}

@misc{tauceti,
  title        = {{Tau Ceti}},
  author       = {{TauCetiProject}},
  howpublished = {\url{https://github.com/TauCetiProject/TauCeti}},
  year         = {2026}
}

@article{collins2024,
  title   = {Evaluating Language Models for Mathematics through Interactions},
  author  = {Collins, Katherine M. and Jiang, Albert Q. and Frieder, Simon and others},
  journal = {Proceedings of the National Academy of Sciences},
  volume  = {121},
  number  = {24},
  pages   = {e2318124121},
  year    = {2024},
  doi     = {10.1073/pnas.2318124121}
}

@article{collins2026,
  title   = {Characterizing Initial Human-{AI} Proof Formalization Workflows},
  author  = {Collins, Katherine M. and Frieder, Simon and Bayer, Jonas and
             Loader, Jacob and Lime, Jeck and Song, Peiyang and Zaiser, Fabian and
             Zhou, Lexin and Li, Shanda and Looi, Sam and Tenenbaum, Joshua B. and
             Bhatt, Umang and Weller, Adrian and Hernandez-Orallo, Jose and
             Freer, Cameron E. and Chen, Valerie and Sucholutsky, Ilia},
  journal = {arXiv preprint arXiv:2606.04273},
  year    = {2026}
}

@article{impactaicodingassistants,
  title   = {The Impact of {AI} Coding Assistants on Software Engineering:
             A Longitudinal Study},
  author  = {Vella, Annie and Blincoe, Kelly},
  journal = {arXiv preprint arXiv:2605.23135},
  year    = {2026}
}

@article{youshallnotpass,
  title   = {You Shall Not Pass! {Where} and Why Developers Draw the Line on
             {AI} Autonomy},
  author  = {Choudhuri, Rudrajit and Bird, Christian and Badea, Carmen and
             Gerosa, Marco and Sarma, Anita},
  journal = {arXiv preprint arXiv:2607.00533},
  year    = {2026}
}

@book{partitions,
  title     = {Integer Partitions},
  author    = {Andrews, George E. and Eriksson, Kimmo},
  year      = {2004},
  publisher = {Cambridge University Press}
}

@article{faithfulnessgap2026,
  title   = {Beyond Compilation: Evaluating Faithful Natural-Language-to-{Lean}
             Statement Formalization},
  author  = {Zhang, Ke and Gallardo Candela, Patricio and Murthy, Sudhir and
             Xie, Yi and Wang, Zhi and Raissi, Maziar},
  journal = {arXiv preprint arXiv:2606.31002},
  year    = {2026}
}

\section*{Appendix}
\appendix

\raggedbottom

\section{Related Work: Human Oversight in Software Engineering}\label{app:sw-related}
Patterns similar to those reported for mathematicians~\citep{collins2024,collins2026} appear in software engineering. A longitudinal study of professional developers finds that AI coding assistants shift work from writing code toward directing, evaluating, and correcting generated code, while flow and cognitive load worsen for some developers over time~\citep{impactaicodingassistants}. A separate study of 448 professional developers finds that developers generally accept AI producing work under human oversight, while willingness to grant greater autonomy decreases for tasks associated with accountability and design~\citep{youshallnotpass}. Lea's design responds to the same findings: the mathematician remains able to inspect and control the agent, and formalization is integrated into the workspace where mathematical writing already happens.

\section{Backbone Details}\label{app:backbone}

\para{Blueprints.}
Blueprints are established practice in Lean formalization: Massot introduced the format~\citep{massot-blueprint}, LeanArchitect~\citep{zhu2026leanarchitect} automates blueprint generation, and LeanMarathon~\citep{zhang2026leanmarathon} uses one to organize autonomous formalization. In Lea the blueprint is additionally shared state across applications, and node completion is never stored as an independent status. It is determined from the latest Lean verification result for the corresponding proof.

\para{Sub-agent roles.}
Each role is a YAML file specifying the prompt, the allowed tools, the turn budget, and optionally the model. Two shipped examples illustrate the range: the \emph{premise-search} role can only read the project, while the \emph{proof-candidate} role can edit a proof and repeatedly check it with Lean. Sub-agent trajectories are preserved in the run history and can be inspected from LeaChat.

\para{Verification.}
AXLE evaluates SafeVerify on 1,000 production proof-verification requests and demonstrates the practical role of such strict verification for untrusted proof submissions~\citep{axle}.

\para{Storage.}
SQLite stores the parts of the history required to reconstruct a session: messages, code steps, verification verdicts, and usage. Each code step references a content-addressed proof snapshot, so identical snapshots are deduplicated. When a proof file changes, the exact snapshot sent to the client is stored with the corresponding code step, and when Lean later returns a verdict, Lea attaches it to that same step; the live canvas and a restored session therefore present the same recorded proof state.

\section{Application Interface Examples}\label{app:screenshots}

\begin{figure}[H]
  \centering
  \includegraphics[width=0.8\textwidth]{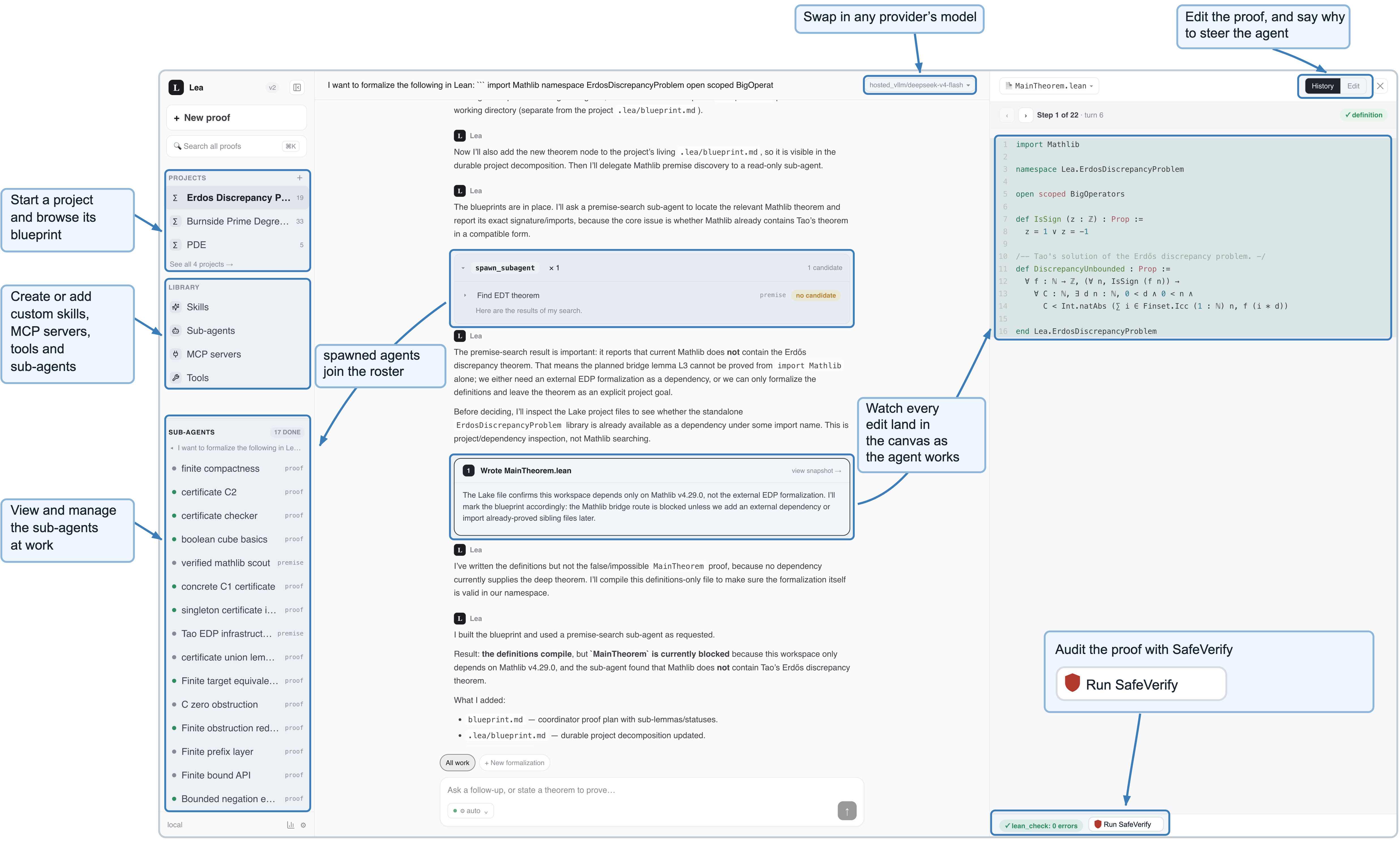}
  \caption{\textbf{The LeaChat interface.} Left: projects and their blueprints, the user-editable library of skills, sub-agent roles, tools, and MCP servers, and the roster of sub-agents spawned in the session. Center: the run transcript, where the agent's reasoning, tool calls, and file writes appear as they happen. Right: the Lean canvas, which shows each edit as it lands, stays editable mid-run, and reports the compile verdict and SafeVerify audit in its status bar. The sidebar composites two scroll positions of one session, since its three sections cannot be on screen together.}
  \label{fig:lea_chat}
\end{figure}

\begin{figure}[H]
  \centering
  \includegraphics[width=\textwidth]{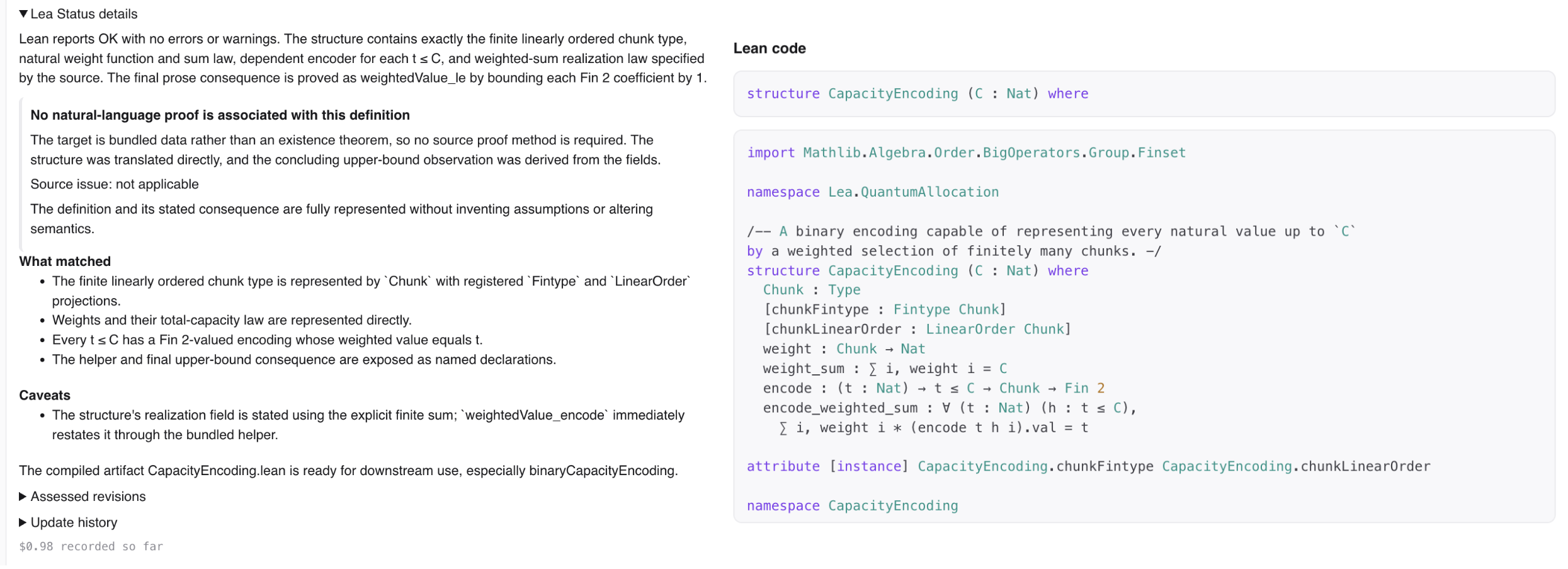}
  \caption{\textbf{LeaOverleaf faithfulness report.} Lea Status report (left) and the corresponding Lean code (right) for a \texttt{CapacityEncoding} definition. The report records the agent's assessment of what in the source was matched, any source issues, and caveats about the formalization. It is a self-assessment, only the mathematician's review can mark the result \emph{audited}.}
  \label{fig:lea_overleaf_faithfulness}
\end{figure}

\section{LeaOverleaf: A Worked Example}\label{app:toy}

An author marks a definition and a theorem that uses it:
\begin{leaexample}
\begin{definition}\label{def:divides}
% lea: define label=divides_def
We say $d$ divides $n$ if $n = dk$ for some $k$.
\end{definition}
\begin{theorem}\label{thm:divisor-bound}
% lea: formalize label=divisor_le
% lea: uses={divides_def}
If $d$ divides $n$, then $d \le n$.
\end{theorem}
\end{leaexample}
The generated \texttt{def} is stated over the integers; the author edits it to $\mathbb{N}$ in the pane (recorded as a human-authored step), re-checks, and audits it. Batch formalization of \texttt{divisor\_le} terminates in \emph{disproved}: every number divides zero. The counterexample identifies a gap in the paper, not the Lean. The author adds $n \ge 1$; the fingerprint change marks the earlier attempt stale, the next run formalizes and verifies the corrected statement, and the author audits it. The definition's audit mark persists, since none of its inputs changed.

\section{Case Study Details}\label{app:casestudy}

The development was produced with GPT-5.5 against Mathlib for Lean 4.32 and consumed 24.7M tokens across three chat sessions and 38 sub-agent invocations. It comprises four core Lean files organized according to the blueprint, including both internal lemmas and outward-facing APIs intended for future work on Ferrers-diagram arguments. Each of the 28 human messages is recorded in the run history with its position relative to agent actions, as is the single sub-agent trajectory that required correction.

\begin{figure}[H]
\label{fig:first_chat}
\begin{leaexample}
You: Begin by defining the Durfee square and Durfee symbol of a partition in Lean.
<@\hrule@>

Lea: Mathematically, for a partition <@$\lambda = (\lambda_1 \ge \lambda_2 \ge \cdots \ge \lambda_\ell > 0)$@>:

...

Does this representation choice match what you want, or would you prefer that I first inspect and use Mathlib's existing partition type directly?
<@\hrule@>

You: Use Mathlib's existing partition type directly.
<@\hrule@>

Lea: Good - we'll use Mathlib's partition type as the source of truth.

...
\end{leaexample}
\caption{\textbf{First Messages.} The mathematician stated each step of the proof outline in plain language and provided short but thoughtful answers to steering questions offered by Lea itself. Being familiar with the subject matter, reviewing Lea's reported design choices and verified mathematical content did not take substantial time for the mathematician.}
\end{figure}

\begin{figure}[H]
\label{fig:bad_agent}
\begin{leaexample}
Lea: The Lean theorem I propose to prove is:

{
    theorem genFun_eq_durfeeCoeff :
        Nat.Partition.genFun = PowerSeries.mk (fun n : ℕ => durfeeCoeff n) := by
}
    
This would live in the Durfee project namespace, probably in a new or existing generating-function file importing Lea.Durfee.DurfeeGenFun and Mathlib's partition generating-function module.

Does this match what you want - namely the power-series statement whose coefficients are the Durfee decomposition counts?
<@\hrule@>

You: This is a valid statement, but not what we're after. We want to use the Durfee decomposition to deduce a different generating function for partitions, namely $$\sum_{k=0}^{\infty} X^{k^2} / (X; X)_k^2$$, which arises from counting the contributions of the Durfee square and symbol to yield all possible partitions.
\end{leaexample}
\caption{\textbf{Significant steering correction.} These messages constitute the largest errant deviation from the proof writing seen during the session, and comprise the longest message sent by the mathematician in order to correct it. The target generating function recounted by the mathematician had been stated in exactly the same form in a prior message.}
\end{figure}

\section{Future Work}\label{app:future}
GitHub integration supports importing and sharing project state but not live collaborative editing. Future work therefore includes collaborative project access, conflict resolution, and per-author provenance, particularly for LeaOverleaf users who already collaborate on the same document. Because Lea records who made each proof change and whether the change passed verification, the run history also supports future studies of when human intervention occurs and how it affects formalization outcomes.

\end{document}